\documentclass[conference]{IEEEtran}
\ifCLASSINFOpdf
\else
\fi

\usepackage{xcolor}
\usepackage{subcaption}
\usepackage{graphicx}
\usepackage{caption}
\usepackage{lipsum}
\usepackage{amsmath,amssymb,amsfonts}
\usepackage[linesnumbered,ruled]{algorithm2e}
\usepackage{algorithmic}
\usepackage{siunitx}
\usepackage{textcomp}
\usepackage{rotating,multirow}
\usepackage{color,soul}
\usepackage{float}
\usepackage{amsmath}
\usepackage{accents}
\usepackage{cite}

\usepackage{booktabs}
\usepackage{graphicx}
\usepackage{multirow}
\definecolor{jmc}{RGB}{199, 0, 57}

    \usepackage[compact]{titlesec}
    \titlespacing{\section}{0pt}{2ex}{1ex}
    \titlespacing{\subsection}{0pt}{1ex}{0ex}
    \titlespacing{\subsubsection}{0pt}{0.5ex}{0ex}

\begin{document}
%

\title{Personalised and Adjustable Interval Type-2 Fuzzy-Based PPG Quality Assessment for the Edge}

\author{


 \IEEEauthorblockN{Jose A. Miranda}
 \IEEEauthorblockA{Embedded Systems Laboratory\\
 \textit{EPFL}\\
 Laussane, Switzerland \\
 0000-0002-7275-4616}

\and
\IEEEauthorblockN{Celia López-Ongil}
\IEEEauthorblockA{Electronic Technology Department \\
\textit{Universidad Carlos III of Madrid}\\
Legan\'es, Spain \\
0000-0001-9451-6611}
\and
\IEEEauthorblockN{Javier Andreu-Perez}
\IEEEauthorblockA{
School of Computer Science \\
and Electronic Engineering \\
\textit{University of Essex}\\
Colchester, UK \\
0000-0002-7421-4808}
%
}

\markboth{Journal of \LaTeX\ Class Files,~Vol.~14, No.~8, August~2021}%
{Shell \MakeLowercase{\textit{et al.}}: ...}

\IEEEpubid{0000--0000/00\$00.00~\copyright~2021 IEEE}

\maketitle

\begin{abstract}
Most of today's wearable technology provides seamless cardiac activity monitoring. Specifically, the vast majority employ Photoplethysmography (PPG) sensors to acquire blood volume pulse information, which is further analysed to extract useful and physiologically related features. Nevertheless, PPG-based signal reliability presents different challenges that strongly affect such data processing. This is mainly related to the fact of PPG morphological wave distortion due to motion artefacts, which can lead to erroneous interpretation of the extracted cardiac-related features. On this basis, in this paper, we propose a novel personalised and adjustable Interval Type-2 Fuzzy Logic System (IT2FLS) for assessing the quality of PPG signals. The proposed system employs a personalised approach to adapt the IT2FLS parameters to the unique characteristics of each individual's PPG signals. Additionally, the system provides adjustable levels of personalisation, allowing healthcare providers to adjust the system to meet specific requirements for different applications. The proposed system 
obtained up to 93.72\% for average accuracy during validation. 
The presented system has the potential to enable ultra-low complexity and real-time PPG quality assessment, improving the accuracy and reliability of PPG-based health monitoring systems at the edge.
\end{abstract}


\begin{IEEEkeywords}
Photoplethysmography, Fuzzy, Personalisation, Signal Quality Assessment, Physiological Monitoring.
\end{IEEEkeywords}

\section{Introduction}
\label{introduction}
\IEEEPARstart{C}urrent physiological monitoring uses wearable and portable sensors to monitor physiological signals like heart rate, blood pressure, and other electrophysiological data, making it more feasible and accessible. This technology can provide continuous and non-invasive monitoring of patients, enabling early detection of abnormalities and timely intervention, ultimately revolutionising healthcare. One common example of continuous physiological monitoring can be found in smart wearables using photoplethysmography (PPG) sensors to track heart rate \cite{ppgReview}. In such cases, while PPG sensors are popular for their cost-effectiveness and ease of integration, they can produce noisy signals due to their optical working principle \cite{PPGSensorComparison}. This creates a challenge for obtaining a clear signal morphology when using wearables. In fact, under real-life conditions, ensuring the robustness and high quality of physiological signals obtained from patients is a challenging task. In this context, Signal Quality Assessment (SQA) systems are responsible for evaluating the quality of acquired signals by analysing different extracted features and applying decision rules to these feature values \cite{bookSOA}. 

When designing SQA systems for physiological monitoring, it's crucial to consider the device's limited computational and memory capabilities and the need for models to generalise across diverse settings. Personalisation is also essential for accuracy and effectiveness in individual patients. However, commercially available monitoring systems prioritise ease-of-use and unobtrusiveness over signal quality, presenting challenges for developing robust and personalised systems \cite{phyChallenges}. Designers must overcome these challenges to create efficient and precise models.

\IEEEpubidadjcol

PPG-SQA systems often overlook important factors such as experimental heterogeneity, generalization, intra- and inter-subject differences, and noise uncertainty. These systems typically rely on hard thresholds and ad-hoc decision rules to determine the signal quality index (SQi) \cite{vadrevu,embedded2,onboardPPG}. However, failing to consider these factors can make it difficult for the system to adapt to new and unseen data. Intra- and inter-subject factors are particularly crucial for ensuring the reliability and robustness of PPG-SQA systems that can work effectively across a diverse population. These factors can significantly impact the morphology of the PPG signal and, consequently, the accuracy of the signal quality assessment.

This paper is based on our previous work presented in \cite{miranda2022wcci} to address these challenges. Based on such research that uses a reduced set of features combined with an interval type-2 fuzzy rule-based system (FRBS) targeting the design of a subject-invariant model, in this case, we propose a late fusion mechanism to consider both a personalised and a global model for the PPG-SQA. This approach deals with the uncertainty principle due to the FRBS and aims to reduce variance and increase generalization due to the joint contribution of the models. Additionally, the system is trained, validated, and tested using our own recollected dataset. 

Combining global models (inter-subject) and local or personalised models (intra-subject) is beneficial for classifying biosignals, as it can reduce performance variances. Global models can account for individual differences across a diverse population and generalise well to unseen data, while local models can capture subject-specific characteristics and stabilise performance. Additionally, the use of local models can reduce the number of required training samples and computational resources, making it more practical for personalised SQA systems. A particular challenge addressed in this paper is enabling the same model to benefit from both global and local perspectives. On this basis, the main objective of this research is to investigate and propose a non-heuristic, edge-oriented and adjustable PPG SQA system by means of a joint contribution of personalised and global FRBS. To the best of our knowledge, this is the first time a late fusion mechanism integrated within an FRBS using personalised and global models is provided.




The rest of the document is structured as follows. 
Section \ref{related} offers an overview of the system framework and discusses recent related work in SQA systems. The proposed system architecture and its data processing pipeline are described in detail in Section \ref{systemDesign}, including the different elements involved in the design. The personalization mechanism employed in the SQA architecture is also explained in Section \ref{systemDesign}. Additionally, Section \ref{experimental_setup} provides an account of the tools employed and database collection. Section \ref{results} showcases the results of the validation and testing phases. Finally, Section \ref{discussion} presents a discussion of the results and a comparison with the state-of-the-art, along with conclusions and future research directions.

\section{Related work}
\label{related}

\subsection{Subject-variability and PPG information}

The inter-subject variability of PPG signals is a major challenge that needs to be addressed for accurate classification and diagnosis. This refers to the differences in PPG signal characteristics between different individuals. This issue has attracted significant attention from the research community, and several studies have been conducted to investigate the inter-subject variability of PPG signals and their impact on different use cases.

One study by Gasparini et al. \cite{gasparini2022} proposes a method for normalising PPG signals based on the individual's resting heart rate, which can vary significantly between individuals. The authors demonstrate that this personalised normalisation method improves the accuracy of PPG-based heart rate variability analysis. Another study by Leitner et al. \cite{Leitner2021} presents a transfer learning approach to estimate blood pressure from PPG signals. The authors train a neural network on a large dataset of PPG and blood pressure measurements from multiple individuals. Then they fine-tune the network on a smaller dataset of PPG signals from a single individual. The results show that the personalised model outperforms a model trained on the entire dataset.

From the analytic perspective, there have also been some proposals. One approach is using normalisation techniques to reduce the impact of inter-subject and intra-subject variability on classification accuracy. For example, Nath et al. \cite{Nath2020} compare two strategies for training a long short-term memory (LSTM) network in classifying electrophysiological signals and showed that the subject-dependent strategy outperforms the subject-independent strategy but also requires more training data. 

\subsection{PPG-SQA State-of-the-art}

PPG-SQA systems can be divided into two different domains based on the features extracted: time and frequency domain. On the one hand, time-domain techniques are commonly used in PPG-SQA systems, as shown in \cite{SQIOptimalTrend}, where the statistical behaviour of different trend-based SQIs was studied. Seven SQIs were tested using 160 recordings, and skewness outperformed the others with an F1-score of up to 87.20\%. However, designing an SQA solely based on these metrics can be prone to heuristic decision rules with hard thresholds. On the other hand, frequency-domain feature extraction-based SQA systems, such as those proposed by Krishnan et al. in \cite{bispectrum}, utilize the spectrum of the skewness of the signal (bi-spectrum) to exploit the phase relations in a clean PPG signal. Different PPG-SQA works can be found based on this taxonomy \cite{vadrevu,embedded2,onboardPPG}. 

Deep and machine learning algorithms can also be used for PPG-SQA systems but are not free from empirically determined decision rules. Moreover, their implementation on edge-computing devices is not as straightforward as the ones solely based on time or frequency features. For instance, in \cite{deep_SQA}, a deep learning model based on convolution neural networks was trained using different time and frequency-domain features, achieving up to 85\% and 83\% for sensitivity and specificity, respectively. However, this approach requires a large memory impact, which is often unfeasible in resource-constrained systems. Alternatively, in \cite{SVM_decission}, pulse wave morphological features were extracted, and different models such as support vector machines and decision trees were trained. The system achieved up to 98.27\% sensitivity using cost-sensitive training. Although the latter is valuable work, the employed dataset was recollected in clinical conditions, which hinders the motion artefacts' heterogeneity. This is an essential factor when considering the possibility of deploying such a system for real-life settings.

In conclusion, inter-subject variability is a major issue in the classification of PPG signals, and several studies have investigated the impact of this issue on classification accuracy. Nevertheless, the solutions proposed so far are based on increasing the complexity of the instrumentation, transforming the signals, or more complex machine learning frameworks. An alternative approach is to use analytical frameworks that intrinsically permit handling uncertainty and imprecision in data. The PPG-SQA can be biased by the effect of inter-subject variability. Still, Fuzzy logic can help to address this problem by allowing for partial membership to different symbolic inputs and classes. This means that a PPG signal can be assigned to multiple concepts and classes with different degrees of membership, reflecting the uncertainty and imprecision inherent in the data. 

The effect of intra-subject variability on SQA can be significant because it can introduce biases into the signal, affecting the accuracy and reliability of any subsequent analysis or classification. Nonetheless, an unresolved inquiry persists regarding the optimal management of subject variability, whereby a PPG-SQA model can simultaneously balance the accommodation of both inter-subject (global) and intra-subject (local or personalised) information. The present study presents a method for reconciling these competing demands.






\section{Personalised and Adjustable PPG-SQA System}
\label{systemDesign}

\begin{figure*}
\centerline{\includegraphics[scale=0.4]{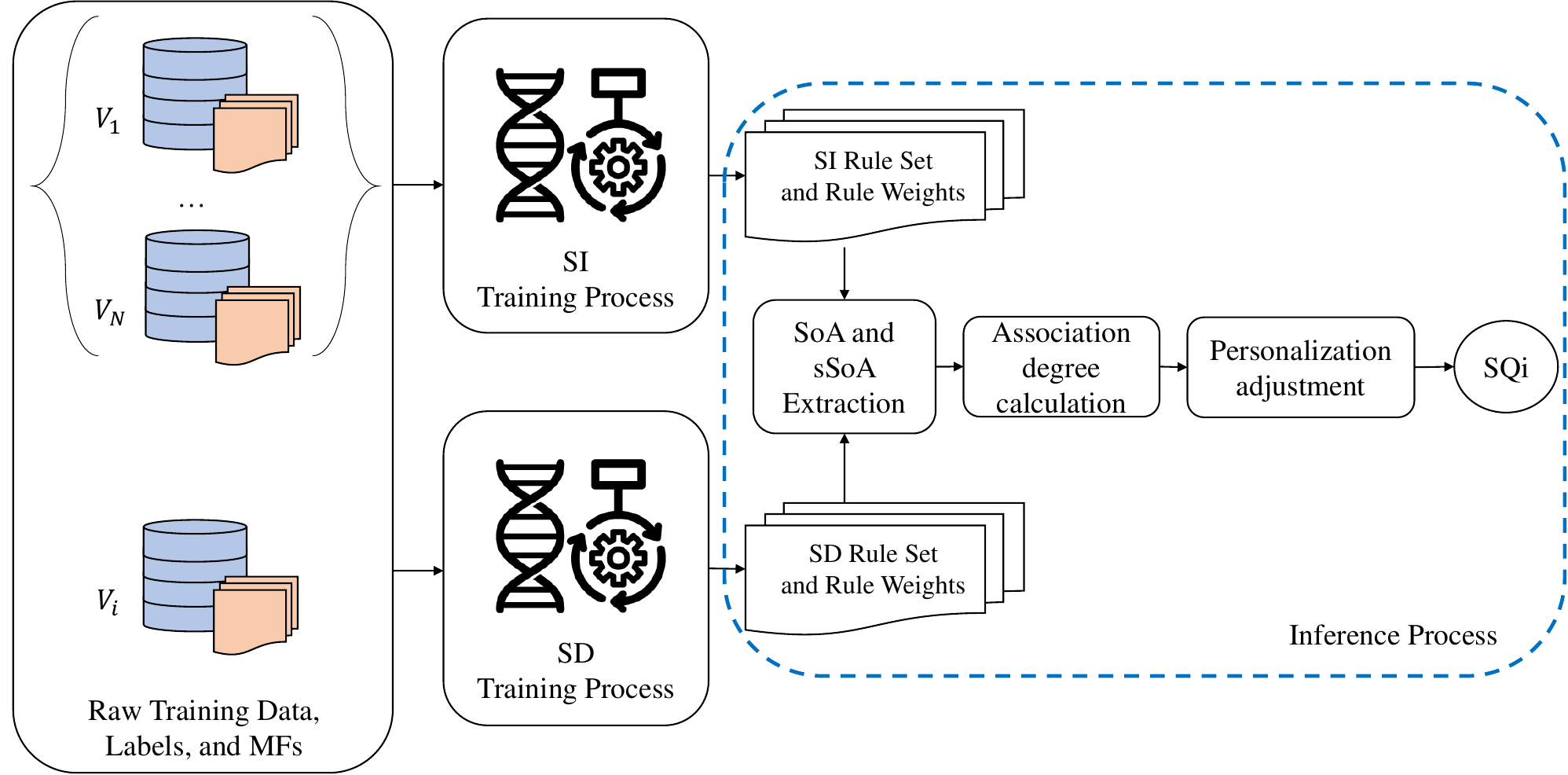}}
\caption{Outline of the proposed architecture for the personalised and adjustable PPG-SQA system.}
\label{figOnline}
\end{figure*}

The main outline of the proposed system architecture is depicted in Figure \ref{figOnline}. Note that the training process is the same as the one presented in \cite{miranda2022wcci}, i.e. the same number of features, quantization method, membership function generation mechanism, and evolutionary genetic algorithm (GA) to optimise the set of rules are also applied for this work. Moreover, the maximum number of antecedents allowed for every rule ($A_{max}$) and the maximum number of total rules ($M$) are also fixed to three and ten, respectively.


One of the main differences with respect to \cite{miranda2022wcci} is that, in this case, two different sets of optimal rules are being generated based on the model, i.e. a Subject-Independent (SI) rule set and a Subject-Dependent (SD) rule set. Note that each rule is expressed by the following nomenclature:
\begin{equation}
R_{M_j}:IF~\phi_a~is~\beta_b~and~...~and~\phi_c~is~\beta_d~then~Y~is~\gamma_{j},
\label{eq_rules_sdsi}
\end{equation}
where $M$ is the specific type of model ($SI$ or $SD$), $a\neq c$, $\phi$ are the different antecedents contained within rule $j$, $\beta$ are the activated linguistic variables for every antecedent, and $\gamma$ is the respective consequent of the rule. Note that, given the different models, two sets of Rule Weights (RW) are generated and assigned to every generated rule for both upper and lower memberships for each of the models. The RW score is calculated as outlined in \cite{antonelli2017}, following:
\begin{equation}
\begin{split}
    &\overline{RW_M}_j=\overline{c_M}_j\cdot\overline{s_M}_j
    \\&\underline{RW_M}_j=\underline{c_M}_j\cdot\underline{s_M}_j
\end{split}
\label{rw_typeII}
\end{equation}
where ${c_M}_j$ and ${s_M}_j$ are the rule confidence and rule support for rule $j$ and model $M$ respectively. The former represents the likelihood of a pattern correctly classifying a sample, while the latter is a quantification of the rule coverage over the training set. 
They can be expressed as
\begin{equation}
\begin{split}
    &c_{M_j}(\phi_j\Rightarrow \gamma) = \frac{\sum_{x_t \in\gamma}w_{M_j}^s(x_t)}{\sum_{j=1}^{T_R} w_{M_j}^s(x_t)}\\
    &s_{M_j}(\phi_j\Rightarrow \gamma) = \frac{\sum_{x_t \in\gamma}w_{M_j}^s(x_t)}{T_R}
\end{split},
\end{equation}
where $x_t$ is every training sample, $T_R$ is the total amount of rules in the rule set, and $w_{M_j}^s$ is the scaled strength of activation of such instance of rule $j$ and model $M$ with input $x_t$. The latter is calculated as follows:
\begin{equation}
    w_{M_j}^s(x_t)=\frac{w_m(x_t)}{\sum_{k,Y=\gamma}w_k(x_t)},
\label{ssoa}
\end{equation}
where $w_m(x_t)$ is the strength of activation, and $w_k(x_t)$ is the sum of all strengths of activation that have the same consequent of rule $j$ for the model $M$. Specifically, the strength of activation is computed as:
\begin{equation}
    w_{M_j}(x_t)=\prod_{z=1}^{A_{max}}\mu_{{\widetilde{A}}}^z(x_t),
\label{soa_typeII}
\end{equation}
where $\mu_{{\widetilde{A}}}^z(x_t)$ represents the membership degree value of the $x_t$ data sample for both the interval type II fuzzy lower and upper membership degree functions.

Finally, the final SQi generation
for each instance is based on the association degree computation with respect to the rule $j$ being evaluated into model $M$. Note that the association degree is given by

\begin{equation}
    \overline{h}_{M_j}(x_t)=\overline{w}_{M_j}^s(x_t)\cdot\overline{RW}_{M_j},  ~
    \underline{h}_{M_j}(x_t)=\underline{w}_{M_j}^s(x_t)\cdot\underline{RW}_{M_j}.
\end{equation}

After that, the overall association degree considering the contribution of the upper and lower type II membership functions for a rule $j$ and a model $M$ is computed as the average between the upper and the lower association degrees. However, in this case, due to the contribution of both sets of rules, we will have an overall SD ($h_{{SD}_j}(x_t)$) and an overall SI ($h_{{SI}_j}(x_t)$) association degree. Once these scores are obtained, the following reasoning or labelling fusion method is employed to assign the predicted class:
\begin{equation}
\begin{split}
    &Y_j = \gamma_j\Rightarrow\\
    &\max_{\forall k\in 
    j}\left(\sum_{k,Y=\gamma_n}(\alpha\cdot h_{{SD}_k}(x_t)+(1-\alpha)\cdot h_{{SI}_k}(x_t)),\right.
    \\&\left.\sum_{k,Y=\gamma_c}(\alpha\cdot h_{{SD}_k}(x_t)+(1-\alpha)\cdot h_{{SI}_k}(x_t))\right),
\end{split}
\end{equation}

where $Y_j$ is the predicted class, and $\alpha$ is the personalisation score that is adjusted during the validation and it can range from $0$ (no personalisation model or personalised rule set contribution) up to $1$ (no global model or global rule set contribution). Note also that, as this is a binary classification problem, we have two consequent: the noisy PPG segment class ($\gamma_n$) and the clean PPG segment class ($\gamma_c$). As far as the author is aware, this is the first instance of an adjustable personalized-based reasoning method being proposed, validated, and tested in an FRBS.



Regarding GA optimisation, particularly for this work, each iteration of the GA algorithm was assessed by conducting k-fold cross-validation on a training-validation split. The split utilized either 3-fold or 5-fold disjoint training and validation datasets, depending on the number of available positive classes and the volunteer. It is important to note that the signal segment acquisition and feature extraction did not undergo any overlapping process. Thus, there was no transfer of information from the learning of rules in one training set or fold to others.




The evaluation of each cross-validated iteration's performance is ultimately measured by calculating the cost using Mathew's Correlation Coefficient (MCC). 
Besides MCC, several other metrics are employed to compare the different cross-validations. These include sensitivity, specificity, geometric mean between sensitivity and specificity (Gmean), and accuracy (ACC).

\section{Tools and Methods}
\label{experimental_setup}

 


We utilized a single dataset for training and validating the proposed SQA. Our own experiment was conducted \cite{miranda2022wemac}, which involved the visualization of diverse audiovisual stimuli through an Oculus Rift\textregistered~ virtual reality headset while recording several physiological variables using wearable sensors, a BiosignalPlux\textregistered~ researcher kit, and Bindi's bracelet \cite{bindi,dcisBVP}. All PPG signals were recorded at 200~Hz. It should be noted that the stimuli were dynamic, allowing volunteers to move freely except for sitting restrictions.

For labelling acceptable and unacceptable PPG segments, an expert familiar with PPG and artefacts manually annotated every non-overlapping 3-second PPG window. Out of the total amount of available volunteers, only $10$ of them were labelled. The training dataset included 331 windows with 269 acceptable and 62 unacceptable PPG segments.

The offline design, implementation, training and validation were done in MATLAB\textregistered~R2021a.








\section{Results}
\label{results}

This section presents the experimental results regarding the validation of the proposed personalised and adjustable PPG-SQA system. 


\subsection{Personalised score exploration}
\label{offvali_personal}

\begin{figure}
\centerline{\includegraphics[scale=0.4]{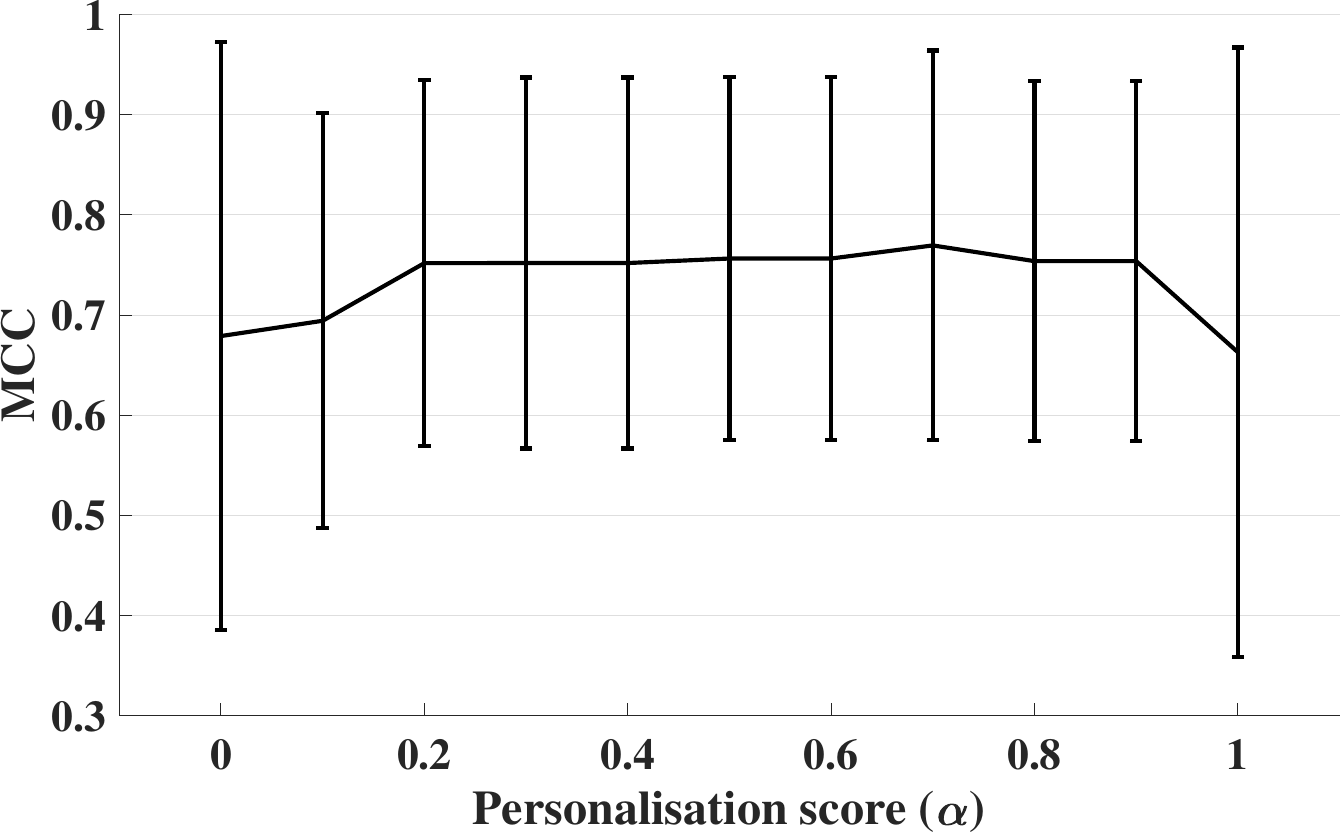}}
\caption{Averaged and standard deviation of the MCC metric vs personalisation score ($\alpha$) sweep, from $0$ to $1$ in $0.1$ steps, for the training.}
\label{figalpha}
\end{figure}

Figure \ref{figalpha} shows the average and standard deviation of the MCC metric for every possible value of $\alpha$ using the training set. Note that the $\alpha$ value is applied once the training is done. Thus, this parameter is applied as a late fusion mechanism to combine the quantitative outputs (association degrees) of both the personalised and the global models. Specifically, we can observe how for both extremes, i.e. fully global without personalisation ($\alpha=0$) and fully personalised without any global contribution, the averaged MCC drops below $0.7$. Moreover, in those use cases, the variance is also affected leading up to an average standard deviation of about $0.3$. This fact proves our previous hypothesis regarding the usage of global and personalised models towards variance reduction. In this case, the best result is obtained for a personalisation score of $0.7$, which provides an averaged MCC of up to $0.77$ with an average standard deviation of $0.19$ for all ten volunteers.

\subsection{Validation: Global vs Personalised model}
\label{offvali}

Table \ref{validationResults} presents the obtained validation results for the global and personalised models. Note that we are presenting the values for the best $\alpha$ concluded after the exploration ($\alpha=0.7$) together with the values for the two extremes to assess and analyse these three model types. 

\begin{table*}[h]
    \centering
    \caption{Validation performance metrics.}
    \begin{tabular}{c c c c c c}
   \toprule 
	 Model  & Sensitivity & Specificity & Gmean & MCC & ACC \\ 
	  Type & $\mu(\sigma)$ & $\mu(\sigma)$ & $\mu(\sigma)$ & $\mu(\sigma)$ & $\mu(\sigma)$ \\
    \midrule
	Global ($\alpha=0$) &  \textbf{90.92 (12.68)} &  {87.11 (21.70)} &  \textbf{88.03 (15.84)} &  {0.68 (0.29)} &  {90.46 (10.13)} \\ 
 	Personalised ($\alpha=1$) &  {80.20 (23.58)} &  {80.89 (38.13)} &  {72.17 (31.71)} &  {0.66 (0.30)} &  {84.57 (26.64)} \\ 
 	Personalised ($\alpha=0.7$) &  {82.70 (24.28)} &  \textbf{91.24 (22.60)} &  {84.19 (18.60)} &  \textbf{0.77 (0.19)} &  \textbf{93.72 (7.44)} \\ 

	\bottomrule
    \end{tabular}
    
    \label{validationResults}
\end{table*}

First of all, we can observe how the sensitivity increases as we decrease the $\alpha$. This behaviour can be due to the fact that the optimised rule set of the global has been obtained using more data, i.e. the amount of noise heterogeneity that the global model knows is higher in comparison to the personalised. On the contrary, the specificity is higher when we combined both models. In fact, the highest value of specificity ($91.24$) is always provided when setting the personalisation score to $0.7$, $0.8$, and $0.9$. Overall, the best result is achieved by the personalised model with $\alpha$ set to $0.7$, which leads up to 0.77\% and 93.72\% of MCC and ACC averaged values, respectively.

\section{Discussion and Conclusion}
\label{discussion}

This study introduces a Mamdani inference model for a fuzzy rule-based system that is personalised, unbiased, and has low complexity. The proposed SQA system has several novel features, including being non-heuristic, adaptive, and personalised. The proposed model's performance is comparable to state-of-the-art methods, achieving an overall validation accuracy of 93.72\%. However, there are limitations to the proposed system. For example, further experimentation and data gathering are required to increase the training data and explore the design space. The advantages and limitations identified during this work highlight the need for SQA systems that can generalise and personalise to heterogeneous settings and volunteers. In forthcoming studies, we could possibly investigate this proposed method in connection with the Internet of Things \cite{vega2022, cortez2020, andreu2009}, other non-invasive physiological sensing modalities \cite{kiani2019}, ambient sensors \cite{andreu2009ambient, andreu2021explainable}, or personal robotics \cite{andreu2018robots}.

% ... 




%
\bibliographystyle{IEEEtran}
\bibliography{main}

\begin{thebibliography}{10}
\providecommand{\url}[1]{#1}
\csname url@samestyle\endcsname
\providecommand{\newblock}{\relax}
\providecommand{\bibinfo}[2]{#2}
\providecommand{\BIBentrySTDinterwordspacing}{\spaceskip=0pt\relax}
\providecommand{\BIBentryALTinterwordstretchfactor}{4}
\providecommand{\BIBentryALTinterwordspacing}{\spaceskip=\fontdimen2\font plus
\BIBentryALTinterwordstretchfactor\fontdimen3\font minus
  \fontdimen4\font\relax}
\providecommand{\BIBforeignlanguage}[2]{{%
\expandafter\ifx\csname l@#1\endcsname\relax
\typeout{** WARNING: IEEEtran.bst: No hyphenation pattern has been}%
\typeout{** loaded for the language `#1'. Using the pattern for}%
\typeout{** the default language instead.}%
\else
\language=\csname l@#1\endcsname
\fi
#2}}
\providecommand{\BIBdecl}{\relax}
\BIBdecl

\bibitem{ppgReview}
D.~Castaneda \emph{et~al.}, ``A review on wearable photoplethysmography sensors
  and their potential future applications in health care,'' \emph{International
  journal of biosensors \& bioelectronics}, vol.~4, no.~4, p. 195, 2018.

\bibitem{PPGSensorComparison}
B.~Bent \emph{et~al.}, ``Investigating sources of inaccuracy in wearable
  optical heart rate sensors,'' \emph{NPJ digital medicine}, vol.~3, no.~1, pp.
  1--9, 2020.

\bibitem{bookSOA}
C.~Orphanidou, \emph{Signal quality assessment in physiological monitoring:
  state of the art and practical considerations}.\hskip 1em plus 0.5em minus
  0.4em\relax Springer, 2017.

\bibitem{phyChallenges}
M.~M. Baig and H.~Gholamhosseini, ``Smart health monitoring systems: an
  overview of design and modeling,'' \emph{Journal of medical systems},
  vol.~37, no.~2, pp. 1--14, 2013.

\bibitem{vadrevu}
S.~Vadrevu and M.~S. Manikandan, ``Real-time ppg signal quality assessment
  system for improving battery life and false alarms,'' \emph{IEEE Transactions
  on Circuits and Systems II: Express Briefs}, vol.~66, no.~11, pp. 1910--1914,
  2019.

\bibitem{embedded2}
G.~Narendra Kumar~Reddy \emph{et~al.}, ``On-device integrated ppg quality
  assessment and sensor disconnection/saturation detection system for iot
  health monitoring,'' \emph{IEEE Transactions on Instrumentation and
  Measurement}, vol.~69, no.~9, pp. 6351--6361, 2020.

\bibitem{onboardPPG}
S.~Alam, R.~Gupta, and K.~D. Sharma, ``On-board signal quality assessment
  guided compression of photoplethysmogram for personal health monitoring,''
  \emph{IEEE TIM}, vol.~70, pp. 1--9, 2021.

\bibitem{miranda2022wcci}
J.~A. Miranda, A.~P. Montoro, C.~L{\'o}pez-Ongil, and J.~Andreu-P{\'e}rez,
  ``Towards interval type-2 fuzzy-based ppg quality assessment for
  physiological monitoring,'' in \emph{2022 IEEE International Conference on
  Fuzzy Systems (FUZZ-IEEE)}.\hskip 1em plus 0.5em minus 0.4em\relax IEEE,
  2022, pp. 1--8.

\bibitem{gasparini2022}
F.~Gasparini, A.~Grossi, M.~Giltri, and S.~Bandini, ``{Personalized PPG
  Normalization Based on Subject Heartbeat in Resting State Condition},''
  \emph{Signals}, vol.~3, no.~2, pp. 249--265, Apr. 2022.

\bibitem{Leitner2021}
J.~Leitner, P.-H. Chiang, and S.~Dey, ``{Personalized Blood Pressure Estimation
  Using Photoplethysmography: A Transfer Learning Approach},'' \emph{IEEE J.
  Biomed. Health Inf.}, vol.~26, no.~1, pp. 218--228, Jun. 2021.

\bibitem{Nath2020}
D.~Nath, Anubhav, M.~Singh, D.~Sethia, D.~Kalra, and S.~Indu, ``{A Comparative
  Study of Subject-Dependent and Subject-Independent Strategies for EEG-Based
  Emotion Recognition using LSTM Network},'' in \emph{{ICCDA 2020: Proceedings
  of the 2020 the 4th International Conference on Compute and Data
  Analysis}}.\hskip 1em plus 0.5em minus 0.4em\relax New York, NY, USA:
  Association for Computing Machinery, Mar. 2020, pp. 142--147.

\bibitem{SQIOptimalTrend}
\BIBentryALTinterwordspacing
M.~Elgendi, ``Optimal signal quality index for photoplethysmogram signals,''
  \emph{Bioengineering}, vol.~3, no.~4, 2016. [Online]. Available:
  \url{https://www.mdpi.com/2306-5354/3/4/21}
\BIBentrySTDinterwordspacing

\bibitem{bispectrum}
R.~Krishnan \emph{et~al.}, ``Two-stage approach for detection and reduction of
  motion artifacts in photoplethysmographic data,'' \emph{IEEE transactions on
  biomedical engineering}, vol.~57, no.~8, pp. 1867--1876, 2010.

\bibitem{deep_SQA}
E.~K. Naeini \emph{et~al.}, ``A real-time ppg quality assessment approach for
  healthcare internet-of-things,'' \emph{Procedia Computer Science}, vol. 151,
  pp. 551--558, 2019.

\bibitem{SVM_decission}
\BIBentryALTinterwordspacing
E.~Sabeti, N.~Reamaroon, M.~Mathis, J.~Gryak, M.~Sjoding, and K.~Najarian,
  ``Signal quality measure for pulsatile physiological signals using
  morphological features: Applications in reliability measure for pulse
  oximetry,'' \emph{Informatics in Medicine Unlocked}, vol.~16, p. 100222,
  2019. [Online]. Available:
  \url{https://www.sciencedirect.com/science/article/pii/S2352914819301856}
\BIBentrySTDinterwordspacing

\bibitem{antonelli2017}
M.~Antonelli \emph{et~al.}, ``Multiobjective evolutionary optimization of
  type-2 fuzzy rule-based systems for financial data classification,''
  \emph{IEEE Transactions on Fuzzy Systems}, vol.~25, no.~2, pp. 249--264,
  2017.

\bibitem{miranda2022wemac}
J.~A. Miranda, E.~Rituerto-Gonz{\'a}lez, L.~Guti{\'e}rrez-Mart{\'\i}n,
  C.~Luis-Mingueza, M.~F. Canabal, A.~R. B{\'a}rcenas, J.~M.
  Lanza-Guti{\'e}rrez, C.~Pel{\'a}ez-Moreno, and C.~L{\'o}pez-Ongil, ``Wemac:
  Women and emotion multi-modal affective computing dataset,'' \emph{arXiv
  preprint arXiv:2203.00456}, 2022.

\bibitem{bindi}
J.~A. Miranda~Calero \emph{et~al.}, ``Embedded emotion recognition within
  cyber-physical systems using physiological signals,'' in \emph{2018
  Conference on Design of Circuits and Integrated Systems (DCIS)}, 2018, pp.
  1--6.

\bibitem{dcisBVP}
J.~A. {Miranda}, M.~F. {Canabal}, L.~{Gutiérrez-Martín}, J.~M.
  {Lanza-Gutiérrez}, and C.~{López-Ongil}, ``A design space exploration for
  heart rate variability in a wearable smart device,'' in \emph{2020 XXXV
  Conference on Design of Circuits and Integrated Systems (DCIS)}, 2020, pp.
  1--6.

\bibitem{vega2022}
C.~F. Vega, J.~Quevedo, E.~Escand{\'o}n, M.~Kiani, W.~Ding, and
  J.~Andreu-Perez, ``Fuzzy temporal convolutional neural networks in p300-based
  brain--computer interface for smart home interaction,'' \emph{Applied Soft
  Computing}, vol. 117, p. 108359, 2022.

\bibitem{cortez2020}
S.~A. Cortez, C.~Flores, and J.~Andreu-Perez, ``A smart home control prototype
  using a p300-based brain--computer interface for post-stroke patients,'' in
  \emph{Proceedings of the 5th Brazilian Technology Symposium: Emerging Trends,
  Issues, and Challenges in the Brazilian Technology, Volume 2}.\hskip 1em plus
  0.5em minus 0.4em\relax Springer, 2020, pp. 131--139.

\bibitem{andreu2009}
J.~Andr{\'e}u, J.~Vi{\'u}dez, and J.~A. Holgado, ``An ambient assisted-living
  architecture based on wireless sensor networks,'' in \emph{3rd Symposium of
  Ubiquitous Computing and Ambient Intelligence 2008}.\hskip 1em plus 0.5em
  minus 0.4em\relax Springer, 2009, pp. 239--248.

\bibitem{kiani2019}
M.~Kiani, J.~Andreu-Perez, H.~Hagras, E.~I. Papageorgiou, M.~Prasad, and C.-T.
  Lin, ``Effective brain connectivity for fnirs with fuzzy cognitive maps in
  neuroergonomics,'' \emph{IEEE Transactions on Cognitive and Developmental
  Systems}, vol.~14, no.~1, pp. 50--63, 2019.

\bibitem{andreu2009ambient}
J.~Andr{\'e}u, J.~Vi{\'u}dez, and J.~A. Holgado, ``An ambient assisted-living
  architecture based on wireless sensor networks,'' in \emph{3rd Symposium of
  Ubiquitous Computing and Ambient Intelligence 2008}.\hskip 1em plus 0.5em
  minus 0.4em\relax Springer, 2009, pp. 239--248.

\bibitem{andreu2021explainable}
J.~Andreu-Perez \emph{et~al.}, ``Explainable artificial intelligence based
  analysis for interpreting infant fnirs data in developmental cognitive
  neuroscience,'' \emph{Communications biology}, vol.~4, no.~1, pp. 1--13,
  2021.

\bibitem{andreu2018robots}
J.~Andreu-Perez, F.~Deligianni, D.~Ravi, and G.-Z. Yang, ``Artificial
  intelligence and robotics,'' \emph{arXiv preprint arXiv:1803.10813}, 2018.

\end{thebibliography}

\vskip -2\baselineskip plus -1fil

\begin{IEEEbiography}[{\includegraphics[width=1in,height=1.25in,clip,keepaspectratio]{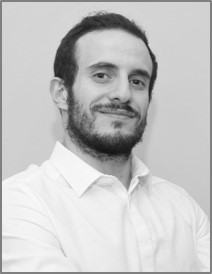}}]{Jose A. Miranda Calero} received the BSc degree in Industrial Electronics and Automation Engineering in 2012, and the MSc in Electronic Systems and Applications Engineering in 2015 with honors, both from Universidad Carlos III de Madrid. He is currently a PhD candidate at the Microelectronic Design and Applications (DMA) research group, which belongs to Universidad Carlos III de Madrid. From 2012 to 2015, he has worked as an Embedded Software Engineer in different countries within Europe for public and private sectors. His research field comprises a wireless sensor, networks, wearable design, development and integration for safety applications, affective computing implementation into edge computing devices, and hardware acceleration. Regarding the development of wearable technology for safety applications, his main contribution relates to the design of BINDI, which is a new autonomous, smart, inconspicuous, connected, edge-computing-based, and wearable solution able to detect and alert when a user is under a gender violence situation employing emotion recognition using the physiological and physical signals of the user. BINDI is being developed within the UC3M4Safety Team. Moreover, he has also participated in other projects related to space applications, such as DS-EXOMARS20.
\end{IEEEbiography}

\vskip -2\baselineskip plus -1fil

\begin{IEEEbiography}[{\includegraphics[width=1in,height=1.25in,clip,keepaspectratio]{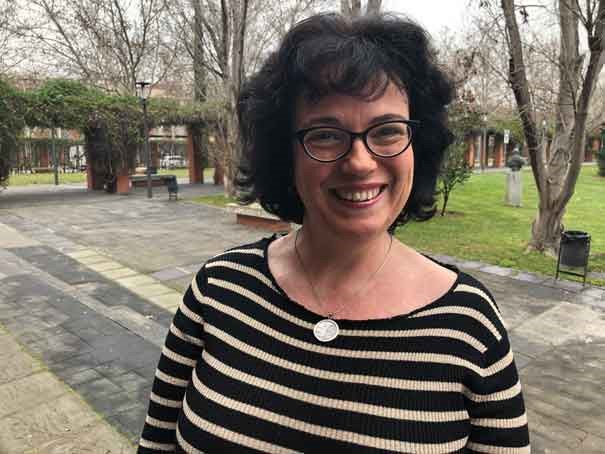}}]{Alba Paez Montoro} received the BSc degree in Industrial Electronics and Automation Engineering in ...
\end{IEEEbiography}

\vskip -2\baselineskip plus -1fil

\begin{IEEEbiography}[{\includegraphics[width=1.05in,height=1.15in,clip,keepaspectratio]{images/celia_paper_sqa.jpg}}]{Celia López-Ongil} is an Industrial Engineer from the Polytechnic University of Madrid (1995) and a PhD in Industrial Engineering from the same university (2000). In 1998 she joined the Universidad Carlos III de Madrid as an assistant professor. She is an Associate Professor at the Department of Electronic Technology since 2010. Her research was framed in the line of Robust Circuits for Space applications in the first 20 years, focused on the generation of HW / SW tools to ensure the quality and reliability of microelectronic circuits, and produced two PhD theses, 19 articles in scientific journals and 90 contributions at international conferences. Some designs in space missions (ASCAT-METOP, SADE-ROSETTA, DS-EXOMARS20) can be highlighted. Since 2016, her research has focused on cyber-physical systems, dedicated to detecting, preventing, and fighting violence against women. She is the leader of the UC3M4Safety multidisciplinary team, that proposes to use technology to protect women at risk of sexual or gender-based violence. The team has been a semi-finalist in the international XPRIZE Women's Safety competition and has won the Vodafone Foundation Award “Connecting for Good” in 2019.
\end{IEEEbiography}

\vskip -2\baselineskip plus -1fil

\begin{IEEEbiography}[{\includegraphics[width=1in,height=1.25in,clip,keepaspectratio]{images/celia_paper_sqa.jpg}}]{Javier Andreu-Pérez} ...
\end{IEEEbiography}

\end{document}